# Effective and efficient ROI-wise visual encoding using an end-to-end CNN regression model and selective optimization

Kai Qiao,[1] Chi Zhang,[1] Jian Chen,[1] Linyuan Wang,[1] Li Tong,[1] Bin Yan[1]

[1] Academy of information systems engineering, PLA strategy support force information engineering university, Zhengzhou, 450001, China.

Correspondence should be addressed to Bin Yan; ybspace@hotmail.com

## Abstract

Recently, visual encoding based on functional magnetic resonance imaging (fMRI) have realized many achievements with the rapid development of deep network computation. Visual encoding model is aimed at predicting brain activity in response to presented image stimuli. Currently, visual encoding is accomplished mainly by firstly extracting image features through convolutional neural network (CNN) model pre-trained on computer vision task, and secondly training a linear regression model to map specific layer of CNN features to each voxel, namely voxel-wise encoding. However, the two-step manner model, essentially, is hard to determine which kind of well features are well linearly matched for beforehand unknown fMRI data with little understanding of human visual representation. Analogizing computer vision mostly related human vision, we proposed the end-to-end convolution regression model (ETECRM) in the region of interest (ROI)-wise manner to accomplish effective and efficient visual encoding. The end-to-end manner was introduced to make the model automatically learn better matching features to improve encoding performance. The ROI-wise manner was used to improve the encoding efficiency for many voxels. In addition, we designed the selective optimization including self-adapting weight learning and weighted correlation loss, noise regularization to avoid interfering of ineffective voxels in ROI-wise encoding. Experiment demonstrated that the proposed model obtained better predicting accuracy than the two-step manner of encoding models. Comparative analysis implied that end-to-end manner and large volume of fMRI data may drive the future development of visual encoding.

## Introduction

In neuroscience, the mechanisms by which visual stimuli are encoded by neurons has yet to be elucidated. Functional magnetic resonance imaging (fMRI)[1] can reflect brain activity effectively, hence, visual encoding models[2, 3] that predict corresponding fMRI signal in response to external visual stimuli, have attracted much too attention in these years. During constructing computation models, new knowledge about the mechanisms can be discovered or validated[4, 5]. In addition, visual encoding is also the base of visual decoding[6-10] that predict stimuli information from fMRI signal, as well as some applications of brain-machine interfaces[11].

In human vision system, the mapping from external visual stimuli to human activity is usually deemed highly nonlinear. In order to stimulate the nonlinear mapping, linearizing encoding manner[3] is the first option and has been used widely in the visual encoding domain, although non-linearizing encoding manner also has started to attract attention[12]. The kind of linearizing encoding method is mainly composed of one nonlinear mapping from visual stimuli space to feature space, and one linear mapping from feature space to brain activity space. The plain linear regression model with specific regularization is used to realize the linear mapping, hence, constructing one encoding model almost focuses on the nonlinear mapping part, namely feature transformation or feature representation[13].

In computer vision domain, how to construct the better nonlinear feature transformation aimed is also the most critical part regardless of image classification or object detection or other task[14, 15]. It is obvious that visual encoding and computer vision domain share a lot in term of feature transformation. Hence, visual encoding models are designed mainly based on existing feature transformation in computer vision. In the early period of computer vision domain, many methods mainly depend on some hand-crafted feature transformation[16], such as gabor wavelet pyramid (GWP)[17], histogram of oriented gradient (HOG)[18], local binary patterns (LBP)[19], scale-invariant feature transform (SIFT)[20], and et al. Kay et al. employed GWP features to construct the famous encoding model[21], and obtained major improvement for encoding primary visual cortices. For high-level visual cortices, visual encoding models are usually based on hand-marked semantic labels[4], because high-level semantic features are hard to design manually. Since the big breakthrough[22] made by deep network and big data[23], deep network with hierarchical feature transformation have driven the enormous advance in computer vision domain[24, 25]. In contrast, deep network can automatically learn or mine effective features from big data for specific task, especially for those task that requires high-level semantic features. Shortly afterwards, the hierarchical and powerful feature transformation was introduced into visual encoding domain[26], and obtained better encoding performance and deeper understanding of human vision system once again[27-30]. Recently, visual encoding has had many achievements with the rapid development of deep network computation, including some new network architectures, such as ResNet[31], recurrent neural network[32], variational autoencoder[33] and capsule network[34]. From the above review, the two domains have been crossing and learning from each other. Researchers have been pursuing a kind of well matching nonlinear feature transformation with human vision system. Besides, how to choose suitable features is important and directly influences encoding performance. Currently, some hand-crafted and learned features have been validated and accepted, for example, Gabor features are fit for visual representation of primary visual cortices, and hierarchical pre-trained convolutional neural network (CNN) architecture better accords with hierarchical visual representation in human vision system. Although these method have obtained fine results, there is still far from accurately predicting human brain activity[8, 30], which suggests that these features still are not optimally matched to brain activity, hence, are not optimal for visual encoding. However, it is hard to determine what kind of feature transformation is the optimal for visual encoding, with only little unclear prior information at hand. These prior information usually confines that primary and high-level visual cortices are responsible for low-level features (edges, corner, and so on) and abstract features (object shape, category, and so on), respectively. Hence, a very natural question emerges for visual encoding domain: *with the development from hand-crafted feature to deep learning features, from computer vision to visual encoding domain, what is next breakthrough for visual encoding?*

In conclusion, current encoding paradigm can be defined as the two-step manner of encoding including firstly choosing well matching feature transformation with corresponding brain activity, and secondly encoding each voxel through linear regression. Essentially, this kind of two-step manner easily fall into local optimal status, and cannot approach to global optimal status. Analogizing computer vision, except CNN architecture, end-to-end manner and big data also play an important role in improving the performance of two-step manner of features engineering (eg. hand-crafted features and classifiers for classic classification task) and have driven enormous advance in computer vision. CNN features have been used to encode voxels, but the employed CNN feature extractor is learned from computer vision task instead of visual encoding task. Hence, they are hard to match with beforehand unknown fMRI data. That is to say, data and task are mismatching[35], which makes it inappropriate to directly transfer fixed CNN feature extractor into visual encoding domain. Inspired by which, we think about abandoning two-step manner of encoding and introducing the end-to-end manner into encoding model. In this way, the encoding model can automatically learn optimally matching features and linear weights (linear regression) from fMRI data for visual encoding.

Previously, the linear regression model used voxel-wise manner[3], and one individual linear regression model needs to be trained for each voxel. Eventually, thousands of regression models are constructed for several visual regions of interest (ROIs). In order to improve the encoding efficiency, we assume that voxels from one visual ROI can be characterized by one kind of specific features and propose a new ROI-wise encoding model that encodes entire voxels in each visual ROI once.

However, not all voxels in one visual ROI can be effectively or easily encoded, which can be seen from encoding results of many previous work. The phenomenon may be caused by two aspect of problem: employed features are not fit for those voxels, and those voxels cannot be effectively encoded because of much low signal to noise ratio (SNR) that makes them unable to reflect the representations of presented stimuli. The first problem can be solved by better matching features learned through end-to-end manner. The second problem indicated that there may exist ineffective voxels in each visual ROI and these ineffective voxels will influence the optimization of effective voxels during the ROI-wise encoding. Hence, the end-to-end encoding manner needs the selective optimization that learns those features related with effective voxels and ignores those ineffective voxels. In order to deal with the problem, we propose to employ a series of operations for selective optimization to construct an encoding model in an end-to-end manner.

In this study, our main contributions are as follows: 1) we analysed current drawbacks of visual encoding in terms of the development of computer vision and visual encoding domain; 2) we introduced the end-to-end manner to learn better matching features to encode voxels for entire voxels of each visual ROI; and 3) we proposed a series of optimization operations to realize selective optimization during ROI-wise encoding in the end-to-end manner.

## Materials and Methods

### Experimental data

The dataset employed in our work was from the previous studies[21, 36]. The dataset had visual stimuli and corresponding fMRI data, consisting of 1750 training samples and 120 testing samples. The detailed information about the visual stimuli and fMRI data could refer to the previous studies[21, 36], and the dataset can be downloaded from http://crcns.org/data-sets/vc/vim-1.

The visual stimuli consisted of sequences of natural photographs, which were mainly obtained from the famous Berkeley Segmentation Dataset[37]. The content of the photographs included animals, buildings, food, humans, indoor scenes, manmade objects, outdoor scenes, and textures. Photographs were converted into greyscale, and downsampled to 500 pixels. The photographs (500 × 500 pixels) presented to subjects were obtained by centrally cropping, masking with a cycle, placing on a grey background, and adding a white square with size of 4 × 4 pixels in the central position. In total, 1870 images (1750 and 120 images for training and testing, respectively.).

Photographs were presented in successive 4s trials. Each trail contained 1s of the photograph presentation with 200ms flashing frequency and 3s of grey background presentation. The corresponding fMRI data was collected when two subjects viewed the photographs and focused on the central white square of the photographs. The experiment of each subject was composed of five scan sessions, and each session had five training runs and two testing runs. 70 different images were presented two times during every training run and 12 different images were presented 13 times during testing run. Images were randomly selected for each run but they were different in every run. Therefore, all 1750 different (5 sessions × 5 runs × 70) images and 120 different (5 sessions × 2 runs × 12) images for training and testing were presented to the subjects.

The 4T INOVA MRI system with a quadrature transmit/receive surface coil was used to acquire fMRI data. The repetition time (TR) was 1s and isotropic voxel size was 2×2×2.5 mm$^3$ in the single-shot gradient EPI sequence. The acquired data was subjected to a series of pre-processing including phase correction, sinc interpolation, motion correction, and co-registration with the anatomical scan. Voxels were assigned to each visual area based on the retinotopic mapping experiment performed in separate session. Voxels in five regions of interest (V1, V2, V3, V4, and LO) from low-level to high-level visual cortices were collected in the dataset.

**The overview of proposed method**

Generally, two-step manner of visual encoding needs two computation models to accomplish the encoding of voxels. As shown in the Figure 1a, the first model is used to map input stimulus space to feature space (S2F model: feature transformation), and the second model is used to map feature space to voxel space (F2V model: regression). Usually, parameter of S2F model is fixed and does not need to be retraining, and only linear weights of F2V regression model need to be trained. The famous GWP model based on Gabor features can be seen in the Figure 1b. CNN features based two-step manner of encoding model only replaces the Gabor features by CNN features, and similarly the parameters of CNN is fixed.

Convolutional neural network (CNN) with powerful feature representation have been widely used in computer vision domain. In this study, we proposed to employ convolutional regression model and train it in the end-to-end manner (ETECRM) in order to avoid the local optimal of two-step manner of visual encoding (see detailed analysis in the Introduction section). As shown in the Figure 1c, the convolution regression model also can be divided into two parts, and front convolutional operations belong to S2F part and the last linear fully connected (fc) layer is F2V part. Most importantly, the ETECRM is trained in the end-to-end manner. End-to-end learning manner is from deep network learning process in which all of the parameters are trained jointly, instead of step by step. For visual encoding, it means that directly learning the mapping from stimulus space to voxel space (S2V model), that is to say, the parameters of S2F part and F2V part in the convolution regression model are learned in the same time from fMRI data samples. In this way, S2F part can approach to better feature transformation and obtain better encoding performance during optimization.

Different from general linear regression with regularization, F2V part or fc layer employed self-adapting regression weights to render voxels can pay more attention to those features well related with themselves. In addition, ROI-wise encoding manner is used to replace the traditional voxel-wise encoding, which can be seen from Figure 1. About the loss, we employed relation (PC) instead of mean square error (MSE) between observed and predicted responses. In order to avoid the interfering of ineffective voxels during ROI-wise encoding, weighted correlation loss and noise regularization are used to accomplish selective optimization. In conclusion, the proposed method is trained in the end-to-end manner and map stimuli to entire voxels of visual ROI once, which realize an effective and efficient visual encoding model.

Next sections introduces the details of the proposed convolution regression model, loss function for the selective optimization, corresponding control models and the quantification method.

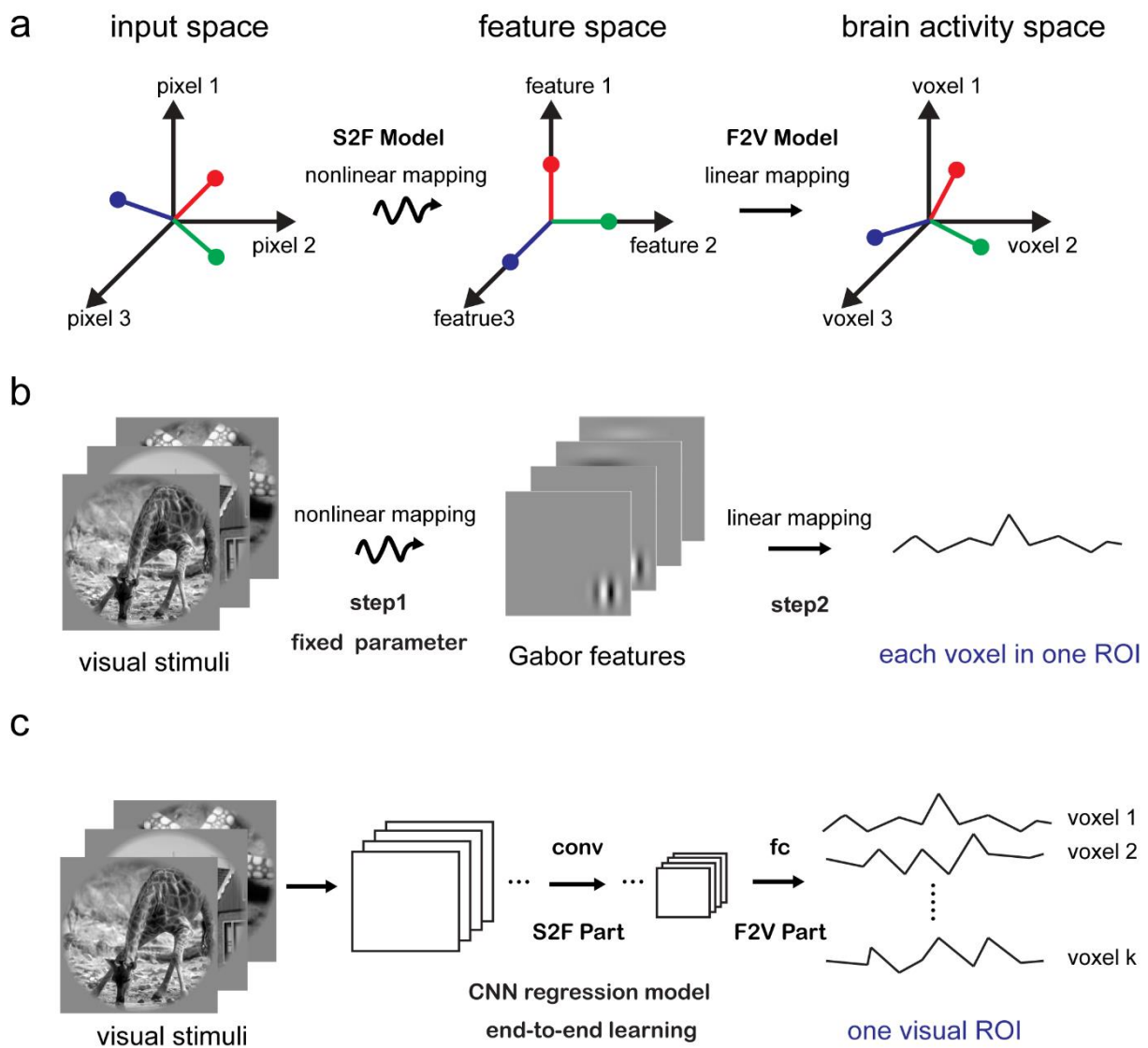

Figure 1. The proposed method including end-to-end learning and ROI-wise encoding. **a**. Three spaces and two mapping are included in the linearizing encoding manner. **b**. Two-step manner of visual encoding including nonlinear feature transformation with fixed parameter and linear

regression mapping. **c**. Convolution regression model for ROI-wise encoding in an end-to-end learning manner.

**CNN feature transformation and self-adapting regression**

In the ETECRM, F2V part mainly includes several convolutional layers and is used to extract features of input image stimuli. Through the stack of convolution filters with setting stride 2 and Rectified Linear Unit (ReLU) activation function, defined feature transformation $F_{w_c}$ in the Equation (1) can transform one image stimulus $s_i$ into convolution features $f_i$, and $w_c$ represents the weights of all convolutional filters in S2V part. As shown in Equation (2), predicted voxels $v'$ in specific visual ROI can be obtained based on one linear regression model. In this study, we assume that bigger weight values in matrix $w_{fc}$ imply that corresponding features are more important than those features whose corresponding weight values are smaller. Hence, we replace $w_{fc}$ with $w_{fc}^2$. In this way, weight learning can dynamically adjust learning rate of weights according current status. Equation (3) give the computation of gradient through the backpropagation[38] with chain rule, hence, learning rate of weights becomes self-adapting $\mu w_{fc}$ from unified learning rate $\mu$ during optimization.

$$f_i = F_{w_c}(s_i) \quad (1)$$

$$v' = w_{fc}f_i + b \rightarrow v' = w_{fc}^2 f_i + b \quad (2)$$

$$\Delta w_{fc} = \mu w_{fc} \Delta v' \quad (3)$$

**Weighted correlation loss function and noise regularization**

In the linear regression mapping, predicted $v'$ represents entire voxels in specific ROI (V1, V2, V3, V4, or LO). Hence, in order to relieve the influence of those ineffective voxels during optimizing effective voxels in the ROI-wise encoding, we add gauss noise $n_g$ with zero mean one variance multiplying by $\varphi$ on the each of predicted voxels $v'$ to restrain those ineffective voxels in the Equation (4). The added noise can make those ineffective voxels harder to optimize from the perspective of SNR. In order to update weights of S2F and F2V parts of ETECRM in and end-to-end manner, we construct one unique loss function. Instead of using MSE, we used PC to measure the predicting performance and Equation (5) presents the computation of correlation for $k_{th}$ voxels in specific visual ROI. Note that, in the experiment, $v_k$ is matrix with dimensionality ($m \times n$). $n$ is the number of voxels in specific ROI, and $m$ represents a batch of samples according to a batch of image stimuli and corresponding CNN features, in this way, each predicted voxel $k$ in one ROI have an PC value defined as $cor_k$.

$$v' = v' + \varphi \cdot n_g \quad (4)$$

$$cor_k = cor(v_k, v'_k) = \frac{Cov(v_k, v'_k)}{\sqrt{Var(v_k) \cdot Var(v'_k)}} \quad (5)$$

Instead of computing the mean value of PC value as loss function for all voxels in entire ROI, we think of introducing the weighted PC loss defined as $L$ to further make the optimization pay attention to those effective voxels. However, whether the $k_{th}$ voxel is important is unknown before optimization, hence the corresponding weight $\eta_k$ is hard to determine in Equation (6).

In this study, we employed current iteration of $cor_k$ to construct corresponding weights. The $cor_k$ ranges from -1 to 1, and the absolute value of $cor_k$ imply the importance of the $k_{\text{th}}$ voxel. In this way, we use the $cor_k^2$ as the weight in Equation (7), and can realize dynamic adjustment during the optimization. In addition, noise regularization is added to prevent the predicted voxels $v'$ from becoming bigger to make the added gauss noise $n_g$ ineffective. Equation (8) gives the final loss function, and the parameter $\gamma$ is used to adjust the proportion of the regular term and fidelity term. In the end-to-end learning, all parameters ($w_c$ and $w_{fc}$) in the ETECRM can be updated at the same time. Equation (9) gives the final optimization problem. We solve the problem through gradient descent based on the open source of deep learning framework PyTorch[39].

$$L = \frac{\sum_k \eta_k cor_k}{n} + \gamma \left|\frac{\sum_k v'_k}{n}\right| \tag{6}$$

$$\eta_k = cor_k^2 \tag{7}$$

$$L = \frac{\sum_k \eta_k cor_k}{n} + \gamma \left|\frac{\sum_k v'_k}{n}\right| = \frac{\sum_k cor_k^3}{n} + \gamma \left|\frac{\sum_k v'_k}{n}\right| \tag{8}$$

$$w' = \underset{w=\{w_c; w_{fc}\}}{\mathrm{argmax}} L = \underset{w=\{w_c; w_{fc}\}}{\mathrm{argmax}} \left(\frac{\sum_i cor_i^3}{n} + \gamma \left|\frac{\sum_i v'_i}{n}\right|\right) \tag{9}$$

**Control models and quantification of encoding performance**

In order to validate the performance of our proposed model, we compare it with two control encoding models. The first model is the classic GWP based model[21] (GWPM). The GWPM is a two-step and voxel-wise encoding model that uses Gabor wavelet pyramid basis functions to construct the feature representation. The mapping from the feature space to the brain activity space is implemented using sparse linear regression[40]. The second model is CNN features based model (CNNM), which is similar to the method described in[28]. This model is a two-step and voxel-wise encoding model and uses the pre-trained CNN features in AlexNet[22] as nonlinear feature extractors to construct the feature space. This linear mapping uses the same sparse linear regression approach as above. Since AlexNet has 8 layers and in theory any layer can be used as the feature space to predict response, we established 8 encoding models from layer 1 to layer 8 respectively and select the model with the highest prediction accuracy as the control model.

We define the prediction accuracy for a voxel as the Pearson correlation (PC) between the observed and the predicted responses across all 120 images in the testing set. For each voxel, we calculated the corresponding prediction accuracy for the three models (GWPM, CNNM, and ETECRM). We firstly made a scatter plot in which each dot corresponds to a single voxel (Figure 4a and 5a). The ordinate value of each dot represents the prediction accuracy of one control model (GWPM or CNNM), while the abscissa value represents the prediction accuracy of the ETECRM. Secondly, we plotted the distribution of prediction accuracy difference of the voxels on whom both models yielded significant prediction. Here, the correlation threshold for significance prediction is 0.27 (p<0.001)[21] (Figure 4b and Figure 5b). Lastly, the voxels in each ROI were sorted in the descending order of the prediction accuracy values (Figure 6), in order to analyse the relationship between the prediction accuracy and the number of effective encoding voxels in the three models.

## Results

**Experiment Details**

The table 1 presents the details of network configuration and Table 2 presents corresponding encoding performance for the two subjects. Because lower-level and higher-level visual ROIs have smaller and bigger receptive fields, respectively, the S2F part used for feature transformation employed smaller convolutional kernel (3×3) for V1 and V2, and bigger convolutional kernel (5×5) for V3, V4, and LO. The F2V part includes one fully connected (fc) layer for linear regression. Super parameter $\varphi$ and $\gamma$ is given in the Table 2. During the end-to-end training, we employ PyTorch deep network framework[39], set batch size as 64, and use Adam optimization (learning rate is 0.001) to update parameters of S2F and F2V part all together through gradient descent. About 20 epochs are required to accomplish the training on the Ubuntu 16.04 system with one NVIDIA Titan Xp graphics card.

Table 1: Network configuration. The "conv" represents the convolutional layer with the kernels "3×3" or "5×5", and "fc" is the fully connected layer used as regression from features to predicted voxels.

| ROI | V1 | V2 | V3 | V4 | LO |
|---|---|---|---|---|---|
| S2F part | 3×conv (3×3) | 3×conv (3×3) | 3×conv (5×5) | 4×conv (5×5) | 4×conv (5×5) |
| F2V part | 1×fc | 1×fc | 1×fc | 1×fc | 1×fc |

Mean Person correlation (PC) of entire voxels and selected effective voxels (Topk, k=300) in each visual ROI are given in Table 2, and detailed comparison with control models can be seen in the below subsection. Figure 2 presents the trend of the predicting accuracy during training an end-to-end encoding model for V1, and demonstrates robust and stable performance through the selective optimization including self-adapting weights, weighted correlation loss and noise regularization. Meanwhile, we can see that many voxels are not successfully encoded, in terms of big difference between mean and Top300 correlation, which validate the necessary of selective optimization in ROI-wise encoding.

Table 2: Encoding performance for the two subjects. "Mean PC" represents the mean Person correlation of entire voxels in specific visual ROI, and "Top 300 PC" represents the mean Person correlation of top 300 voxels in specific visual ROI.

| Subject | - | | V1 | V2 | V3 | V4 | LO |
|---|---|---|---|---|---|---|---|
| Subject 1 | | $\varphi$ | 1.0 | 1.0 | 1.0 | 1.0 | 1.0 |
| | | $\gamma$ | 1e-3 | 1e-3 | 1e-4 | 1e-5 | 1e-5 |
| | Mean PC | | 0.257 | 0.189 | 0.110 | 0.085 | 0.046 |
| | Top300 PC | | 0.650 | 0.612 | 0.429 | 0.312 | 0.131 |
| Subject 2 | | $\varphi$ | 0.1 | 0.1 | 0.1 | 0.1 | 0.1 |
| | | $\gamma$ | 1e-4 | 1e-4 | 1e-5 | 1e-5 | 1e-5 |
| | Mean PC | | 0.162 | 0.129 | 0.067 | 0.043 | 0.029 |
| | Top300 PC | | 0.491 | 0.442 | 0.248 | 0.052 | 0.031 |

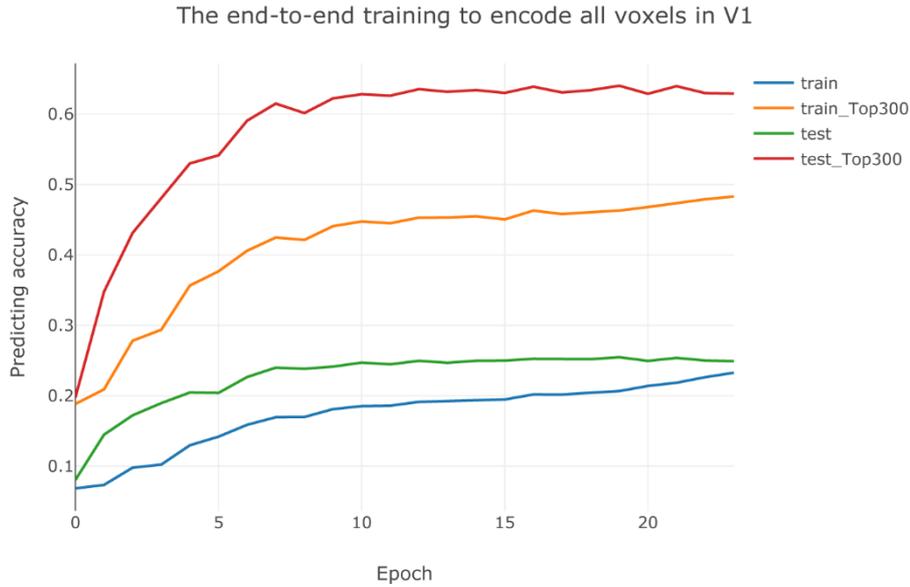

Figure 2: The trend of prediction accuracy on training set and testing set during training encoding model for V1. The stable performance indicate the robustness of end-to-end training, and "Top300" represents the mean correlation of selected voxels with top 300 predicting accuracy.

**Selective optimization**

A series of optimization strategies including self-adapting regression weights, weighted correlation loss and noise regularization are used in the proposed ETECRM in order to selectively optimize those effective voxels and suppress those ineffective voxels. Figure 3 presents the distribution of prediction accuracy of V1 area (1294 voxels in all) on training set and testing set, respectively. We can see the sparse distribution of predicting correlation, and voxels have higher or lower predicting correlation. By restraining those ineffective voxels and paying more attention to those important voxels, the model can automatically learn effective features suitable for important voxels from fMRI data in an end-to-end manner, and obtain better encoding performance (See Figure 4 and 5). The sparse distribution validate the selective optimization of proposed method during ROI-wise encoding.

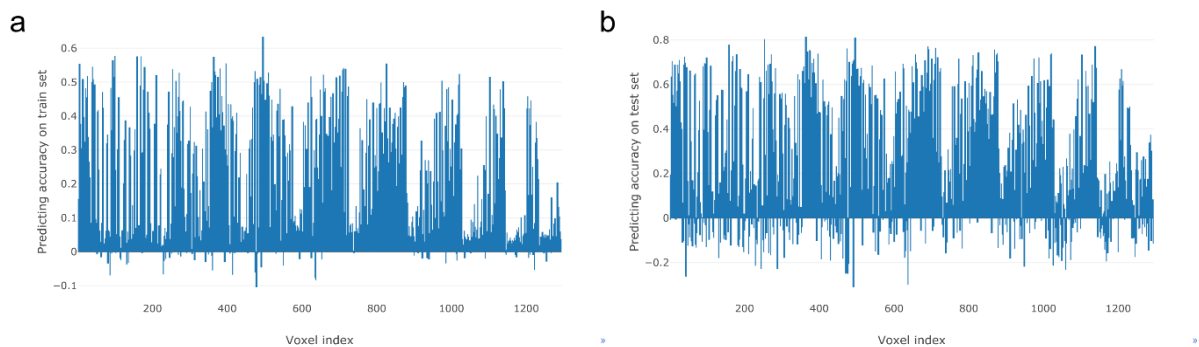

Figure 3: The distribution of prediction correlation of each voxels in V1 area on (a) training set and (b) testing set during optimization, respectively.

**Comparison with two-step manner of visual encoding**

Firstly, we compare with GWPM in terms of prediction accuracy in Figure 4. It can be seen that the ETECRM model was significantly better than the GWPM in all ROIs. From V1 to LO, the percentage of those voxels that can be better encoded by our model are about 90%, which demonstrates significant advantage of end-to-end learning than hand-crafted Gabor features. The ETECRM almost exceeds all voxels, especially in V4 and LO, there are only a few voxels whose responses can be explained by the GWPM and the ETECRM model almost performed better in all voxels. Note that we selected subject 1 to demonstrate the results and analysis, and consistent results for subject 2 can be seen in the Appendices A.

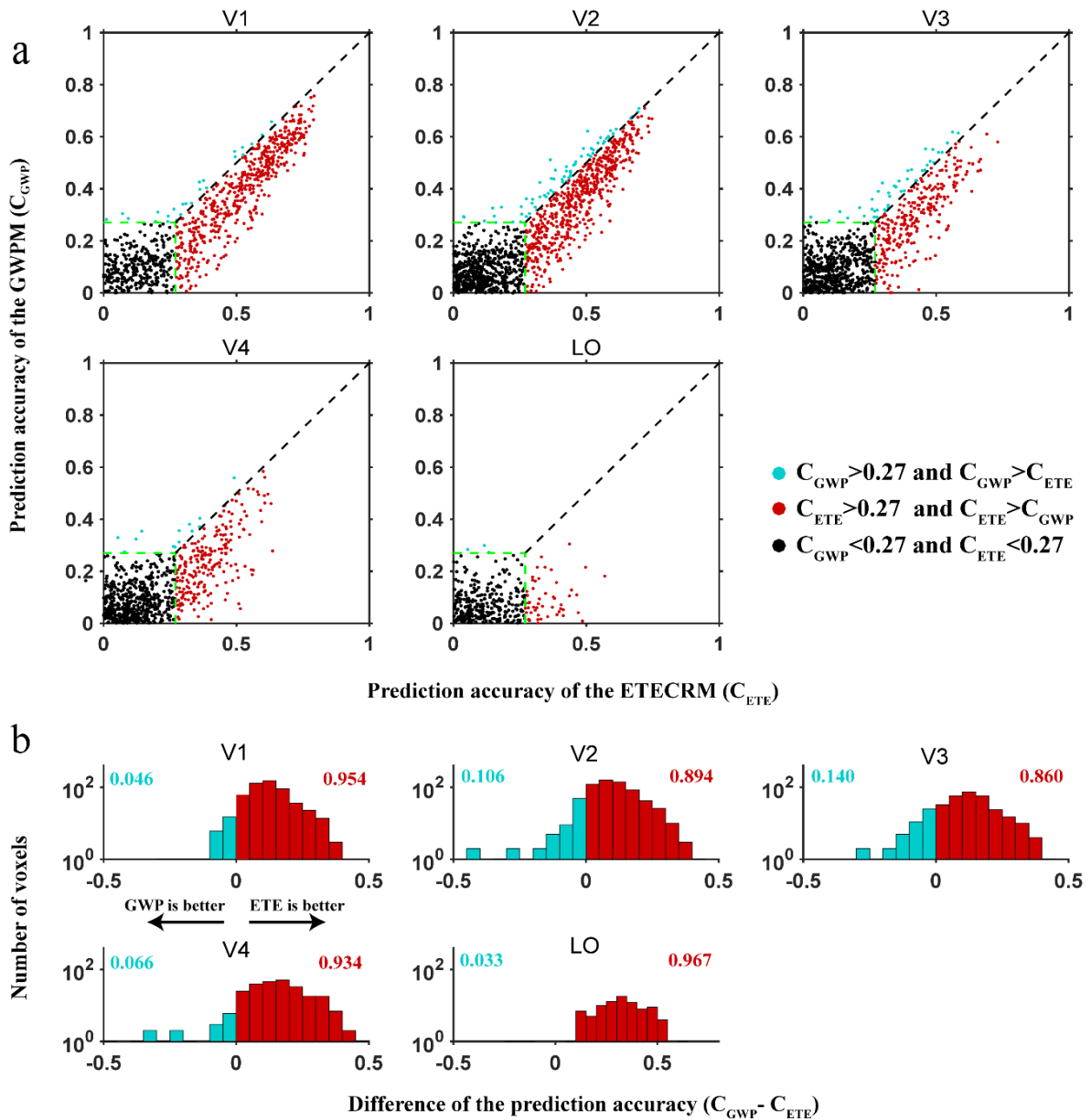

Figure 4: Comparison of prediction accuracy between the ETECRM and GWPM. (a) Each of the five axes displays a comparison between the prediction accuracy of the two models in specific visual ROI. In all five scatter plots, the ordinate and abscissa represent the prediction accuracy values of GWPM and ETECRM respectively. The green dashed lines indicate the significant prediction value of 0.27 (p < 0.001). The black dots indicate the voxels cannot be significantly encoded (under 0.27) by either of the two models. The red dots

indicate the voxels that can be better predicted by the ETECRM than the GWPM and vice versa for the cyan dots. (b) Distribution of the prediction accuracy difference between the ETECRM and GWPM. Prediction accuracy difference above 0 indicates higher prediction accuracy of the ETECRM, as marked by red colour, and vice versa for the cyan colour of the GWPM. The number on each side represents the fraction of voxels whose prediction accuracy are higher under that model.

Next, we compare with the CNNM in terms of prediction accuracy in Figure 5. The ETECRM has higher accuracy than the CNNM in V1, V2, and V3, and V4, and slightly worse in LO. Especially for V1, about 80% voxels can be better encoded by the ETECRM. From V1 to LO, we can see that the advantage of the proposed ETECRM becomes less and less (Figure 5b), which indicates that higher-level feature representation matched with higher-level visual ROI become harder to learn in the end-to-end manner, and pre-trained CNN is trained from millions of samples that demonstrate the advantage in the encoding of high-level visual areas.

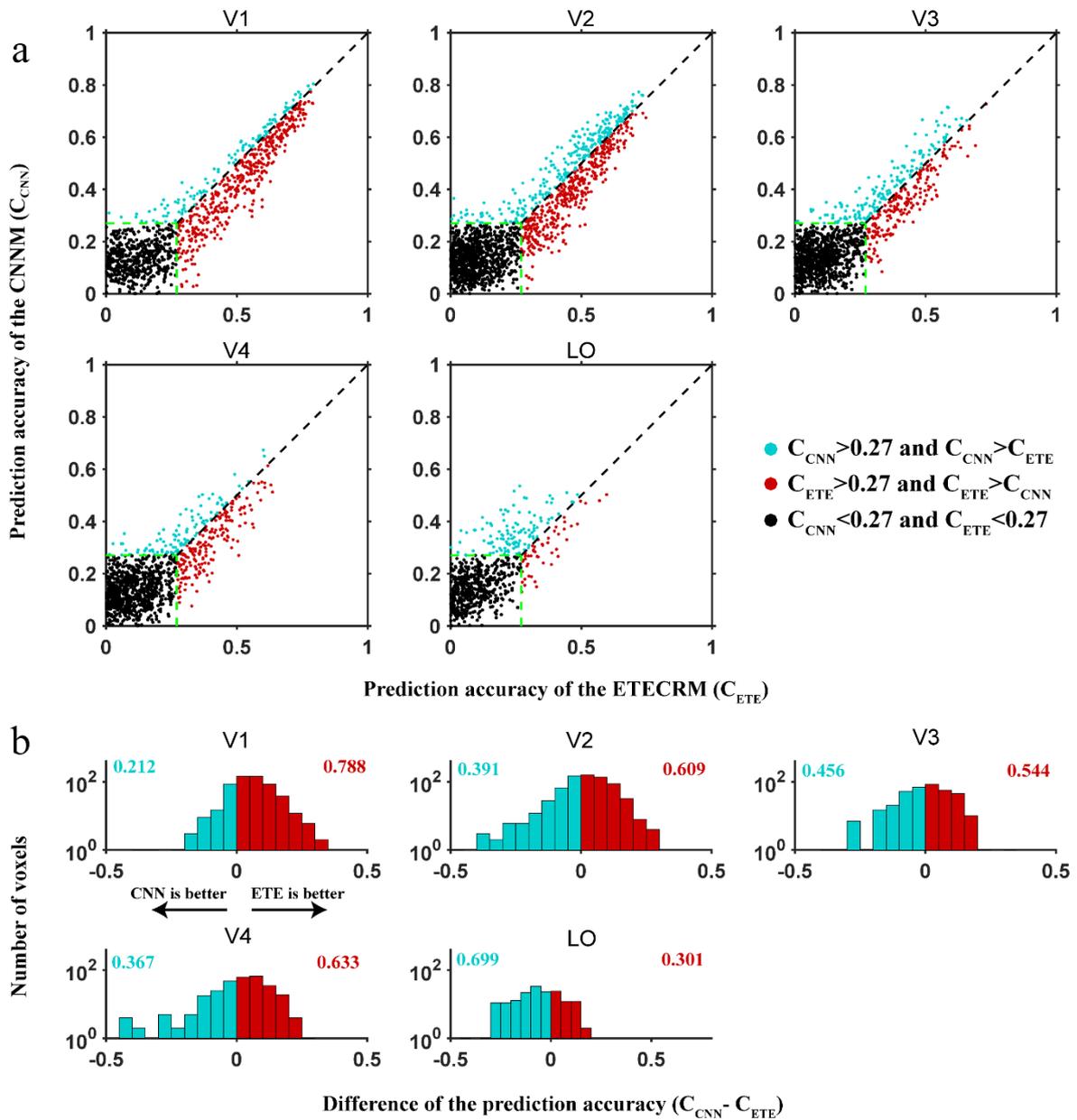

Figure 5: Comparison of prediction accuracy between the CNNM and the ETECRM. All definitions in the subplots are the same as Figure s, except that here the control model is the CNNM.

**Model comparison by sorting voxels in prediction accuracy**

In addition, we extracted those voxels whose responses can be significantly predicted by each of the three models (PC > 0.27) and sorted them in a descending order of the prediction accuracy in the Figure 6. Overall, the ETECRM (red lines in Figure 6) and the CNNM (blue lines in Figure 5) are obviously better than the GWPM (cyan lines in Figure 6). Hence, we mainly focus on comparing the ETECRM and the CNNM. The ETECRM model could significantly predict responses of 41.1%, 31.1%, 17.3%, 16.2% and 9.48% voxels in V1, V2, V3, V4, and LO, respectively. In V1, V2, and V4, the prediction accuracy of the top voxels calculated by the ETECRM is higher than that of CNNM. In V3, similar performance was obtained by the ETECRM and CNNM. In LO, the ETECRM behaves worse than CNNM. We speculate that relatively more data is required for encoding high-level visual ROIs that are responsible for high-level visual representation, and current less data suppresses the function of end-to-end manner, hence, reduce the encoding performance.

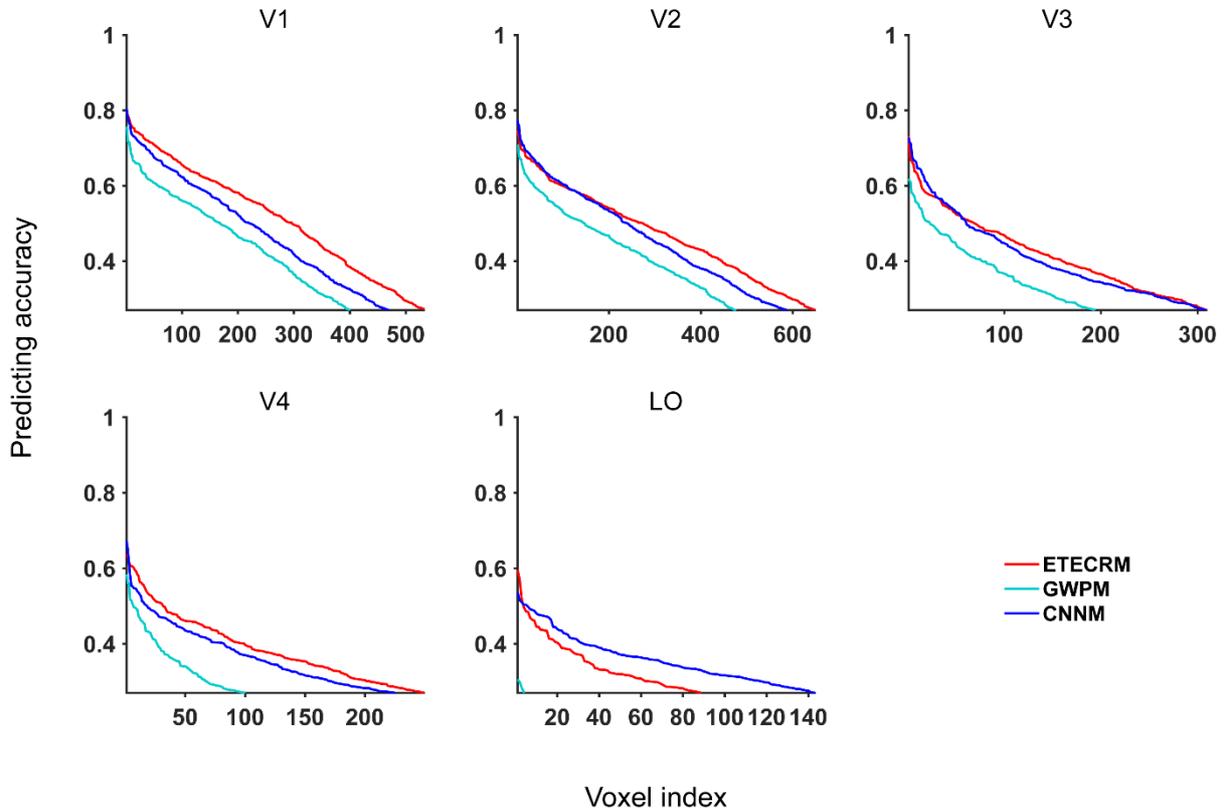

Figure 6. Comparisons of the ETECRM, GWPM, and CNNM by sorting voxels in the descending order of prediction accuracy. Only those voxels on whom the prediction accuracy value exceeds 0.27 are plotted. Red, cyan and blue curves represent the ETECRM, the GWPM, and the CNNM, respectively. The CNNCRM exhibits significantly higher performance in lower level areas, such as V1 and V2.

## Discussion

**Performance of the proposed method**

In this study, we proposed an ROI-wise visual encoding using an end-to-end CNN regression model. Through selective optimization, we successfully trained the model, and obtained higher encoding performance than the two-step manner of models (GWP and CNN features based models). From the perspective of effectiveness, the proposed method avoid the local optimal of two-step manner of encoding, and automatically learn better matching features with brain activity. From the perspective of efficiency, the proposed method accomplished the ROI-wise encoding with better performance. In conclusion, effective and efficient encoding is accomplished based on the proposed method.

**End-to–end manner on visual decoding**

Visual decoding that decodes information (category or presented image stimuli) according to the acquired fMRI data, compared to with visual encoding. Visual decoding is usually accomplished in one inverse two-step manner. Similarly, specific feature space are required as middle bridge that connected voxels and stimuli. However, whether to be able to find matching features also limits the performance of decoding. In this way, the end-to-end manner that can learn the best matching features from fMRI data still can contribute to the decoding domain, and make it easy to reconstruct images based on voxels, regions, and even all visual cortices. Hence, the end-to-end manner may drive the advance of encoding and decoding.

**How to better encode high-level visual ROIs**

Regardless of two-step manner of encoding or end-to-end manner of encoding, the encoding performance of high-level visual ROIs such as LO are worse compared to the encoding of low-level visual ROIs. High-level visual ROIs are responsible for complex semantic visual representations, which are hard to characterize. In this way, trying seeking matching features with those voxels in computer vision domain seems hard to realize. In contrast, the proposed method still seems helpless. However, the encoding results have validated the advantage of end-to-end manner, and we think that complex high-level ROIs can be encoded with more fMRI data, namely the potential of end-to-end manner was restricted by current limited number of fMRI data, compared to big data that have million samples in computer vision. End-to-end manner have leaded the rapid development with big data at hand in computer vision, inspired by which, collecting more data will be the next direction for visual encoding domain. Although minority of researchers [35]start to be aware of the problem of data, their volume of data still cannot be called "big data", comparing to classic ImageNet in the computer vision domain. Visual encoding can refer to the development path of computer vision domain, and the encoding domain might be taking off by combining end-to-end manner and big fMRI data.

## Conclusions

Traditional the two-step manner of encoding including firstly choosing well matching feature transformation and secondly encoding voxels through linear regression in voxel-wise manner. However, it is hard to determine well matched features, and easily falls into local optimal status, and cannot approach to global optimal status. The two-step and voxel-wise manner restrict the encoding effectiveness and efficiency, respectively. In this study, we proposed the ETECRM for ROI-wise encoding and designed the selective optimization to further improve the ROI-wise encoding performance. In this way, we successfully realized one effective and efficient encoding model, and presented better encoding performance for V1, V2, and V3. In addition, one referable way to develop computation neuroscience model from perspective of computer vision was provided and further we give rise to consideration of potential of end-to-end manner and large volume of fMRI data for the future visual encoding.

## Data Availability

The detailed information about the fMRI data is provided in previous studies[21, 36], and the public dataset can be downloaded from *http://crcns.org/data-sets/vc/vim-1*.

## Conflicts of Interest

The authors declare that there is no conflict of interest regarding the publication of this paper.

## Funding Statement

This work was supported by the National Key Research and Development Plan of China (No. 2017YFB1002502), the National Natural Science Foundation of China (No. 61701089), and the Natural Science Foundation of Henan Province of China (No. 162300410333).

## Appendices A. Results for subject 2

The results of subjects 2 were consistent with those of subjects 1. Figure. A1, A2, and A3 correspond to the section "Comparison with two-step manner of visual encoding" in the main text.

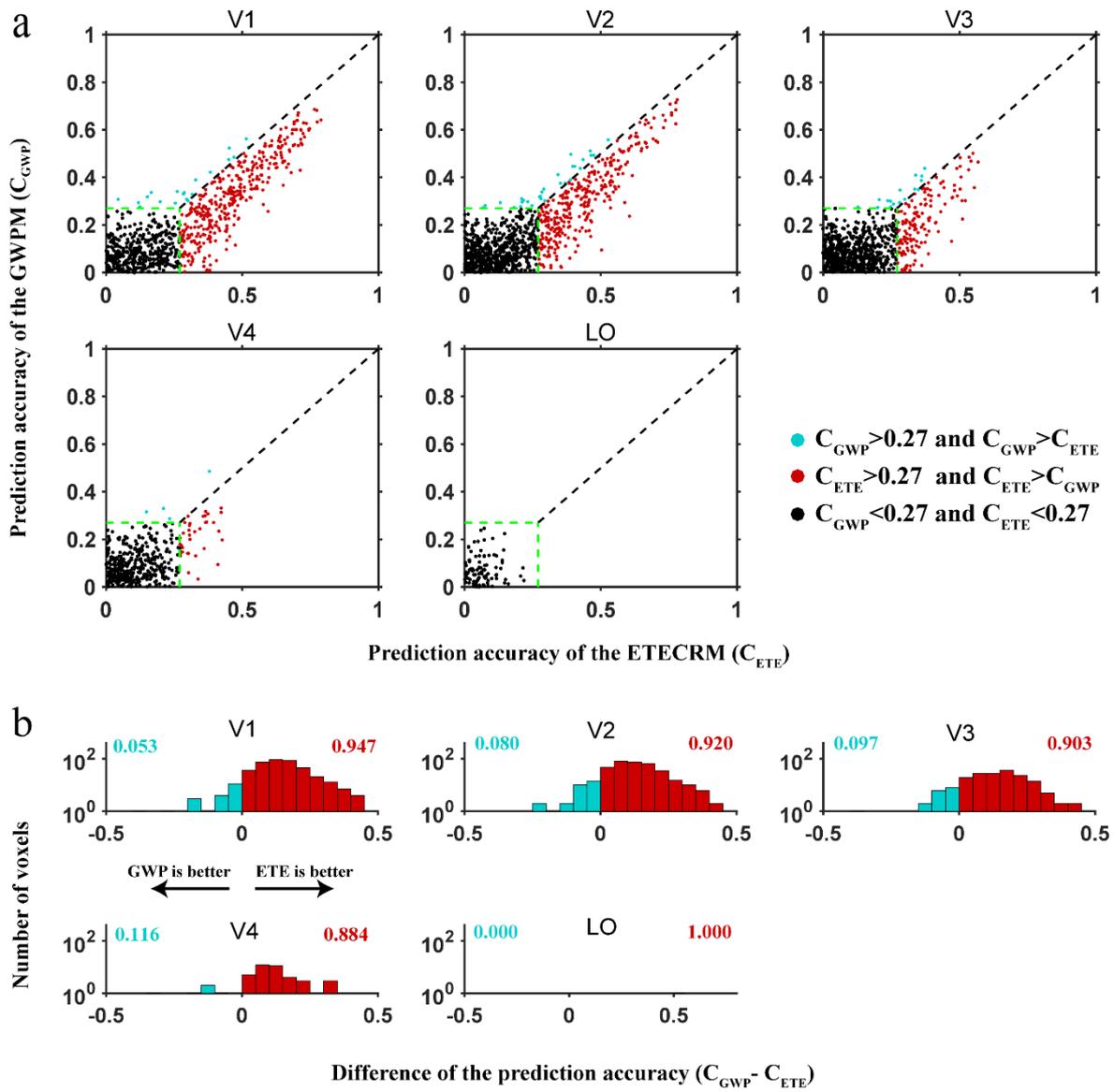

Figure A.1. Comparison of prediction accuracy between the ETECRM and GWPM for subject 2. Refer to Figure 4 for a detailed description of the plot elements.

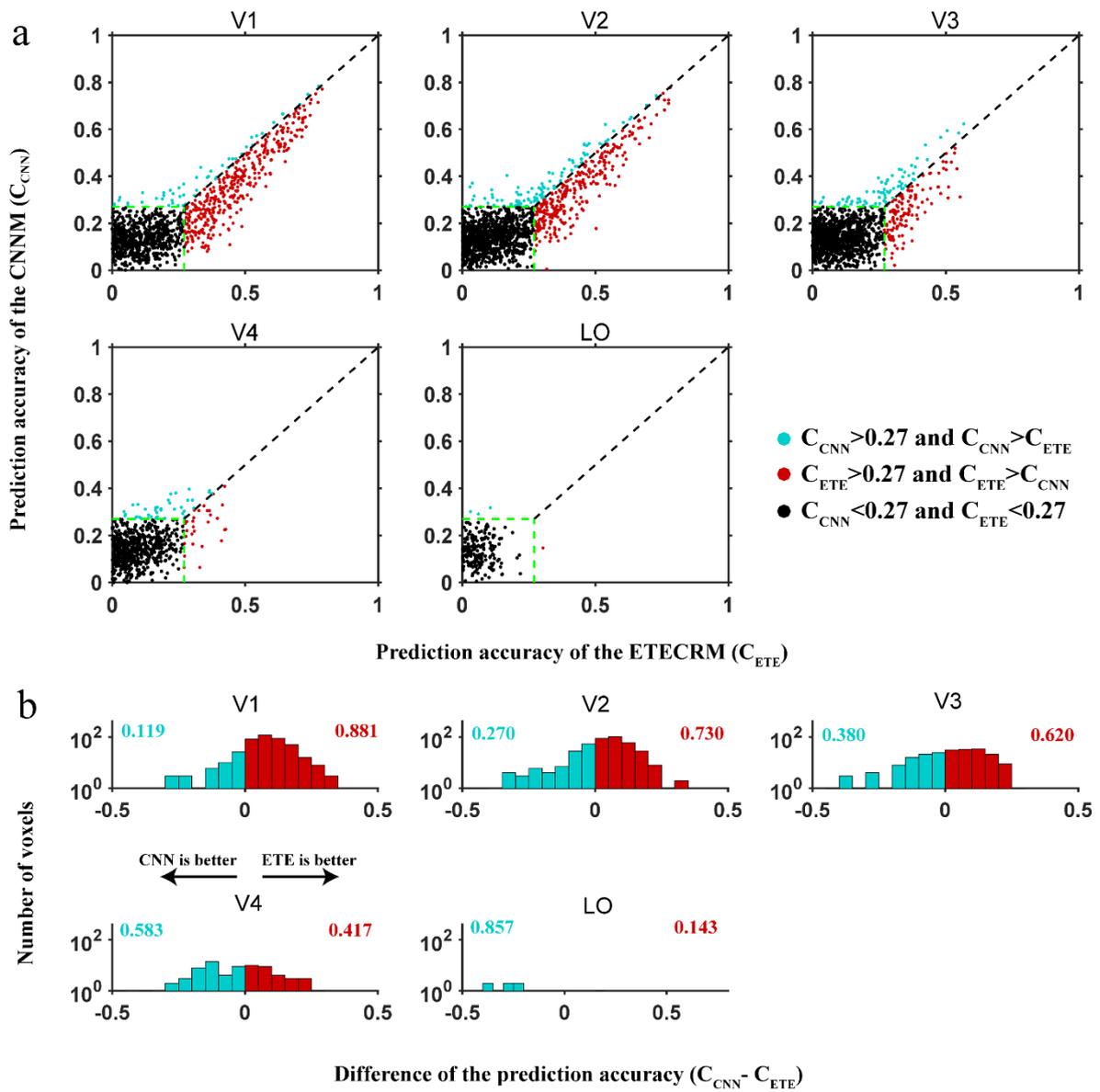

Figure A.2. Comparison of prediction accuracy between the ETECRM and GWPM for subject 2. Refer to Figure 5 for a detailed description of the plot elements.

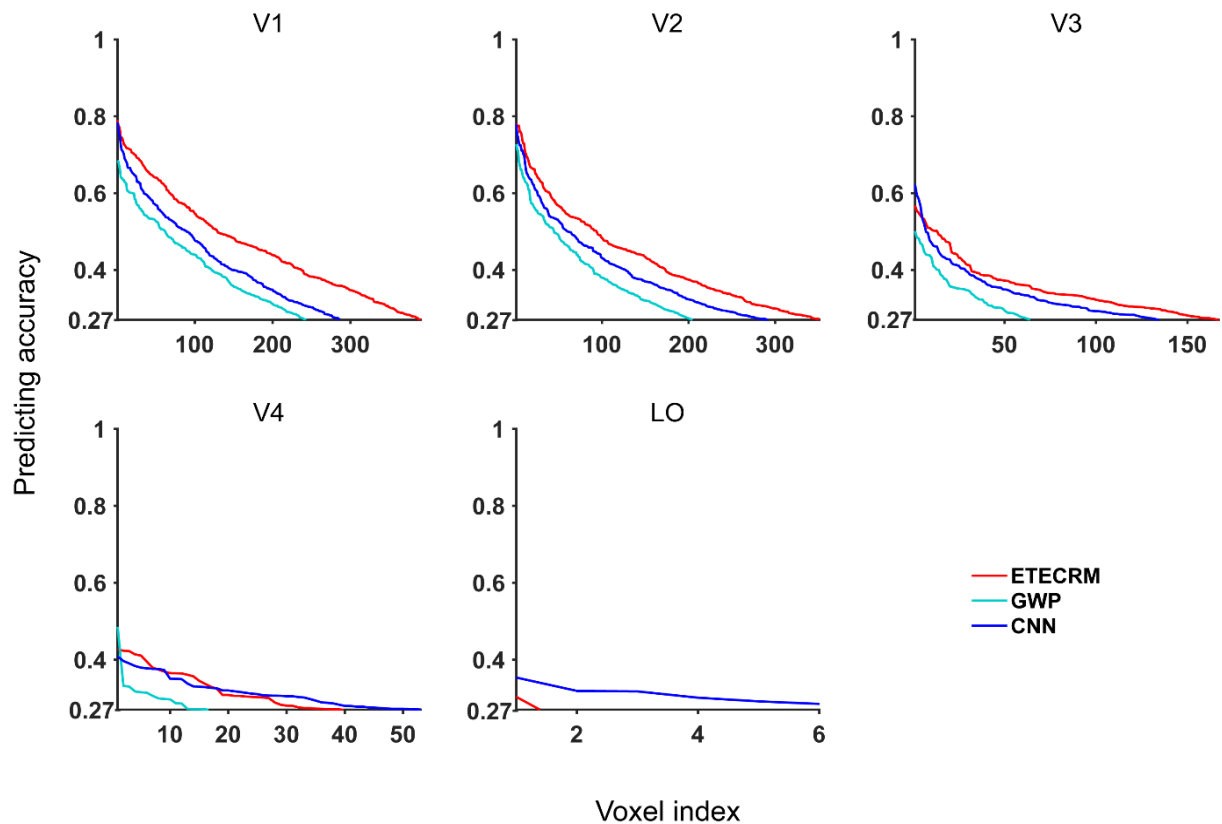

Figure A.3. Comparisons of the ETECRM, GWPM, and CNNM by sorting voxels in the descending order of prediction accuracy for subject 2. Refer to Figure 6 for a detailed description of the plot elements.